\documentclass[preprint,aps,floats,nofootinbib]{revtex4} 
\usepackage{graphicx} 
\newcommand{\be}{\begin{equation}} 
\newcommand{\ee}{\end{equation}} 
\newcommand{\gsim}{\, \raisebox{-0.8ex}{$\stackrel{\textstyle >}{\sim}$ }} 
\newcommand{\lsim}{\, \, \raisebox{-0.8ex}{$\stackrel{\textstyle <}{\sim}$ }} 
 
\newcommand{\feyn}[1]{{#1}\!\!\!{\slash}}

\newcommand{\roughly}[1]%  
{\mathrel{\raise.4ex\hbox{$#1$\kern-.75em\lower1ex\hbox{$\sim$}}}}  

\newcommand\beq{\begin{eqnarray}}  
\newcommand\eeq{\end{eqnarray}}

\def\Dsl{\,\raise.15ex \hbox{/}\mkern-12.8mu D}

\def\fm3{fm$^{-3}$}

\def\ni{\noindent} 
 
\begin{document} 
%\begin{frontmatter} 
% 
\preprint{\vbox{\hbox{SUNY-NTG-02-05-25, MIT-CTP-3269}}} 
 
\bigskip 
\bigskip

\title{Color-Neutral Superconducting Quark Matter} 
\author{Andrew W. Steiner$^1$,   
Sanjay Reddy$^2$, and  
Madappa Prakash$^1$ 
} 
\address{$^1$ Department of Physics and Astronomy,  
State University of Stony Brook, NY 11794 \\ 
$^2$Center for Theoretical Physics, Massachusetts  
Institute of Technology, Cambridge, MA 02139 }

\begin{abstract} 
We investigate the consequences of enforcing local color neutrality on 
the color superconducting phases of quark matter by utilizing the 
Nambu-Jona-Lasinio model supplemented by diquark and the t'Hooft 
six-fermion interactions.  In neutrino free matter at zero 
temperature, color neutrality guarantees that the number densities of 
$u,~d,~{\rm and}~s$ quarks in the Color-Flavor-Locked (CFL) phase will 
be equal even with physical current quark masses.  Electric charge 
neutrality follows as a consequence and without the presence of 
electrons.  In contrast, electric charge neutrality in the less 
symmetric 2-flavor superconducting (2SC) phase with $ud$ pairing 
requires more electrons than the normal quark phase.  The free energy 
density cost of enforcing color and electric charge neutrality in the 
CFL phase is lower than that in the 2SC phase, which favors the 
formation of the CFL phase.  With increasing temperature and neutrino 
content, an unlocking transition occurs from the CFL phase to the 2SC 
phase with the order of the transition depending on the temperature, 
the quark and lepton number chemical potentials.  The astrophysical 
implications of this rich structure in the phase diagram, including 
estimates of the effects from Goldstone bosons in the CFL phase, are 
discussed. 
 
%\bigskip 
%\noindent PACS:   
 
\end{abstract} 
\maketitle 
 
\newpage 
 
Studies of QCD at high baryon density have led to the expectation that 
quark matter is a color superconductor in which the pairing gaps of 
unlike quarks ($ud,~us,~{\rm and}~ds)$ are as large as 100 MeV. For 
three massless flavors, a symmetric ground state called the 
Color-Flavor-Locked (CFL) phase, in which BCS-like pairing involves 
all nine quarks, is favored \cite{Alford:1999mk,Alford99}.  At lower 
density and for physically relevant values of the strange current 
quark mass ($100 < m_s/{\rm MeV} < 300$), a less symmetric (2SC) phase 
in which only the light up and down quarks ($m_{u,d} \le 10$ MeV) pair 
is expected \cite{Alford:1997zt,Rapp:1997zu}. For recent reviews, see 
Refs.~\cite{Rajagopal:2000wf}. 
 
With the exception of the work by Iida and Baym \cite{Iida:2000ha}, 
and more recently by Alford and Rajagopal \cite{Alford:2002kj}, little 
attention has been paid to the issue of color neutrality in 
superconducting quark phases. These works are the primary motivation 
for this study. The issues addressed in this work are similar to those 
addressed by Alford and Rajagopal\cite{Alford:2002kj} who perform a 
model independent analysis that is valid when $m_s \ll \mu$ and 
$\Delta \sim m_s^2/\mu$, where $\Delta$ is the pairing gap and $\mu$ 
is the quark number chemical potential. We employ an extended version 
of the the Nambu-Jona-Lasinio model (NJL hereafter), which shares many 
symmetries with QCD including the spontaneous breaking of chiral 
symmetry, and calculate the thermodynamic potentials, $\Omega$, and 
pairing gaps, $\Delta$, self-consistently in the CFL and 2SC phases. 
Our analysis leads to results that complement some of the conclusions 
in Ref.~\cite{Alford:2002kj}. There are, however, several aspects in 
which we go further. First, we employ a self-consistent model which 
uniquely determines both the diquark and the quark-anti quark 
condensates. Second, since the realization of color and electric 
charge neutrality becomes non-trivial only for physically relevant 
values of $m_s$, we retain terms to all orders in $m_s$ in our 
calculation of $\Omega$. As noted in Ref.\cite{Alford:2002kj}, this is 
particularly important for understanding the phase structure of quark 
matter at densities (or equivalently, $\mu$) of relevance to neutron 
stars, since $m_s/\mu$ is not small compared to unity.  In addition, 
we establish the phase structure of superconducting color-neutral 
quark matter at finite temperature and lepton content which was not 
considered in \cite{Iida:2000ha,Alford:2002kj}, but is relevant for 
studies of proto-neutron stars. 
 
\ni {\bf Charges, chemical potentials, and color neutrality: } Bulk, 
homogeneous matter must be neutral with respect to charges which 
interact through the exchange of massless gauge bosons.  Otherwise, 
the free energy density cost would be infinite.  In the CFL phase, 
diquark condensation breaks color symmetry and all eight gluons become 
massive via the Higgs mechanism. Similarly, in the 2SC phase, 
$SU(3)_c$ is broken down to $SU(2)_c$ and five of the eight gluons 
become massive. In both the CFL and 2SC phases, however, an $U(1)$ 
gauge symmetry remains unbroken\cite{Alford:1999mk}. The associated 
charge is called $\tilde{Q}$. The generator for this charge in the CFL 
phase is a linear combination of the usual electric charge $Q$ and a 
combination of color generators $T_3$ and $T_8$, and is given by  
 
\be 
\tilde{Q}=Q-\frac{1}{2}~T_3-\frac{1}{2\sqrt{3}}~T_8\,, \ee  
 
\ni where $Q=diag(2/3,-1/3,-1/3)$ in flavor space, and 
$T_3=diag(1,-1,0)$ and $T_8=diag(1/\sqrt{3},1/\sqrt{3},-2/\sqrt{3})$ 
in color space. 
 
The color superconducting phase is, by construction, neutral with 
respect to $\tilde{Q}$ charge. Why then should we impose, in addition, 
local color neutrality? As noted earlier, gluons become massive in the 
superconducting phase and the free energy density cost of realizing a 
non-zero color density in bulk matter need not be infinite. Further, 
although a finite sample embedded in the normal state must be a color 
singlet, this alone does not require local color neutrality since 
color singletness is a global constraint. Hence a heterogeneous phase 
with colored domains of typical size similar to the color Debye 
screening length is a possibility. However, in a homogeneous and color 
conducting medium a color charge density in the bulk is unstable as it 
generates a chromo-electric field resulting in the flow of color 
charges~\cite{Alford:2002kj}. Color neutrality is therefore a 
requirement for the homogeneous phase. Neutrality with respect to 
charges associated with $T_3$ and $T_8$ is achieved by introducing 
appropriate chemical potentials $\mu_3$ and $\mu_8$ in analogy with 
the charge chemical potential $\mu_Q$. As noted in 
Refs.~\cite{Alford:2002kj,Amore:2001uf}, color neutrality is a 
prerequisite for color singletness, but the additional free energy 
density cost involved in projecting out the color singlet state is 
negligible for large samples. 
 
The superconducting ground state breaks both color and electromagnetic 
gauge symmetries.  It would therefore seem that excitations above the 
condensate can only be characterized by the unbroken $\tilde{Q}$ 
charge.  At first sight, this would imply that electrons and unpaired 
quarks carry only $\tilde{Q}$ charge and must therefore be assigned 
only a $\mu_{\tilde{Q}}$ chemical potential.  If this were indeed the 
case, it would be impossible to neutralize the 2SC phase in the bulk. 
This is because the condensate is $\tilde{Q}$ neutral, but has color 
and electric charge that cannot be neutralized by particles with only 
$\tilde{Q}$ charge.  The resolution to this puzzle lies in noting that 
our expectation to assign only those charges that are unbroken by 
condensation to excitations applies only to excitations above a 
charge-neutral ground state. In this case, charges associated with 
broken gauge symmetries are easily delocalized and transported to the 
surface by the condensate. It is however important to note that only 
the excess broken charge resides on the surface. In describing 
particles that make up the charge neutral ground state we must use 
vacuum quantum numbers.  In this case, the individual charges are 
localized on the particles in the bulk.  Therefore, in what follows we 
treat electrons and unpaired quarks as carrying their vacuum charges 
in our description of the neutral ground state.

\ni {\bf Thermodynamics:} We begin with the NJL Lagrangian 
\cite{Nambu61,Rehberg96,Buballa99,Steiner00,Buballa01,Gastineau01} 
supplemented by both a diquark interaction and the t'Hooft six-fermion 
interaction which reproduces the anomalous U$_{\mathrm A}$(1) symmetry 
breaking present in QCD \cite{tHooft86}.  Explicitly, 
\begin{eqnarray} 
{\cal L} &=& \bar{q}_{i \alpha} 
\left( i \feyn{\partial} \delta_{i j} \delta_{\alpha \beta} -  
m_{i j} \delta_{\alpha \beta} - \mu_{i j,~\alpha \beta} \gamma^0 
\right) q_{j \beta}  + 
G_S \sum_{a=0}^8 \left[  
\left( \bar{q} \lambda^a_f q \right)^2 + 
\left( \bar{q} i \gamma_5 \lambda^a_f q \right)^2  
\right] \nonumber \\ 
&& - G_{D} \left[  
{\mathrm det}_{i j} \, \bar{q}_{i \alpha} \left( 1 + i \gamma_5 \right)  
 q_{j \beta} + 
{\mathrm det}_{i j} \, \bar{q}_{i \alpha} \left( 1 - i \gamma_5 \right)  
 q_{j \beta} \right] \delta_{\alpha \beta} \nonumber \\ 
&& + G_{DIQ} \sum_k \sum_{\gamma} \left[ 
\left(\bar{q}_{i \alpha} \epsilon_{i j k}  
\epsilon_{\alpha \beta \gamma} q^C_{j \beta}\right) 
\left(\bar{q}_{i^{\prime} \alpha^{\prime}}^C  
\epsilon_{i^{\prime} j^{\prime} k} \epsilon_{\alpha^{\prime}  
\beta^{\prime} \gamma} q_{j^{\prime} \beta^{\prime}}\right)\right.\nonumber \\ 
&& \qquad\qquad\qquad 
 + \left. 
\left(\bar{q}_{i \alpha} i \gamma_5 \epsilon_{i j k}  
\epsilon_{\alpha \beta \gamma} q^C_{j \beta}\right) 
\left(\bar{q}_{i^{\prime} \alpha^{\prime}}^C i \gamma_5  
\epsilon_{i^{\prime} j^{\prime} k} \epsilon_{\alpha^{\prime}  
\beta^{\prime} \gamma} q_{j^{\prime} \beta^{\prime}}\right) \right] \,,  
\label{Lagr} 
\end{eqnarray} 
where $m_{i j}$ is the diagonal current quark matrix, and the spinor $q^C= C 
\bar{q}^T$, where $C$ is the Dirac charge conjugation matrix.  We 
use $\alpha,\beta,\gamma$ for color ($r$= red, $b$= blue, and 
$g$= green) indices, and $i,j,k$ for flavor ($u$= up, $d$= down, and 
$s$= strange) indices throughout. The chemical potential matrix  
is diagonal in flavor and color, and is 
given by 
\begin{equation} 
\mu_{i j,~\alpha \beta} = (\mu \delta_{i j} +  
Q_{i j} \mu_Q ) \delta_{\alpha \beta} + \delta_{i j} \left( T_{3 \alpha \beta} 
 \mu_3 + T_{8 \alpha \beta} \mu_8 \right) \,, 
\label{mui} 
\end{equation} 
\ni where $\mu$ is the quark number chemical potential.  Since the 
couplings $G_S,~G_D$, and $G_{DIQ}$ are dimensionful, we impose an 
ultra-violet three-momentum cutoff, $\Lambda$, and results are 
considered meaningful only if the quark Fermi momenta are well below 
this cutoff.  The values of the couplings are fixed by reproducing 
the experimental vacuum values of $f_{\pi}$, $m_{\pi}$, $m_K$, and 
$m_{\eta^{\prime}}$ as in Ref. \cite{Rehberg96}. For the most part, we  
discuss results obtained using  
\begin{eqnarray} 
m_{0 u} &=& m_{0 d}=5~{\rm MeV}\,, \qquad m_{0 s}=140~{\rm MeV}\,,   
\qquad \Lambda=600~{\rm MeV}\,, \nonumber \\ 
G_S\Lambda^2 &=& 1.84\,, \qquad G_D\Lambda^5=12.4\,, \qquad  
{\rm and} \qquad  G_{DIQ}=3 G_S/4 \,.   
\end{eqnarray}  
In vacuum, the effective four-fermion interactions in the 
$qq$ and ${\bar q}q$ channels are related by a Fierz transformation, hence  
the choice of $G_{DIQ}=3G_S/4$.  
Note that Eq.~(\ref{Lagr}) does not 
include the possible presence of a six-fermion interaction due to 
diquark $(\langle qq \rangle)$ condensates; such interactions have 
been assumed to result only in a renormalization of the four-fermion 
diquark interaction.  In the mean field approximation, the 
thermodynamic potential per unit volume is given by 
\begin{eqnarray} 
\Omega & = &  
- 2 G_S \sum_{i=u,d,s} \langle {\bar q_i} q_i \rangle^2   
+4 G_D \left<{\bar u} u\right> \left<{\bar d} d \right>  
\left<{\bar s} s\right> 
 + \sum_k \sum_{\gamma} \frac{\left|\Delta^{k \gamma}\right|^2}{4 G_{DIQ}} 
\nonumber \\ 
&& - \frac{1}{2} \int \frac{d^3 
p}{\left(2 \pi\right)^3} \, \sum_{i=1}^{72} \left[ \frac{\lambda_i}{2} + 
T \ln{\left(1 + e^{-\lambda_i/T} \right)} \right] +  
\Omega_0 \,, 
\label{Omega1} 
\end{eqnarray} 
where $\langle {\bar q_i} q_i \rangle$  ($i=u,d,s$) is the quark condensate,   
and the term $\Omega_0$ ensures that the zero density  
pressure, $P=-\Omega$, of non-superconducting matter is zero: 
\begin{equation} 
\Omega_0 =   
2 G_S \sum_{i=u,d,s} \langle {\bar q_i} q_i \rangle_0^2  
- 4 G_D \left<{\bar u} u\right>_0 \left<{\bar d} d \right>_0  
\left<{\bar s} s\right>_0 + 2 N_c \sum_i \int \frac{d^3 p} 
{\left( 2 \pi^3 \right) } \sqrt{m_i^2 + p^2} \,, 
\end{equation} 
where $<{\bar q_i} q_i>_0$ denotes the value of the quark condensate at 
zero density.  The gap matrix  
\begin{equation} 
\Delta^{k \gamma} = 2 G_{DIQ} 
\left< \bar{q}_{i \alpha} i \gamma_5 \varepsilon^{i j k} 
\varepsilon^{\alpha \beta \gamma} q^C_{j \beta} \right>  
\label{cfdelta} 
\end{equation} 
features three non-vanishing  
elements. Using the standard notation of denoting $\Delta^{k \gamma}$ 
through the flavor indices $i$ and $j$, we have 
\begin{eqnarray} 
\Delta_{ds} \equiv \Delta^{ur}, \qquad 
\Delta_{us} \equiv \Delta^{dg}, \qquad {\rm and} \qquad 
\Delta_{ud} \equiv \Delta^{sb} \,. 
\label{cfldelta} 
\end{eqnarray} 
This corresponds to the ansatz in Ref. \cite{Alford99}, except that 
color sextet gaps (symmetric in both color and flavor) are ignored. 
Inclusion of the sextet gaps modifies our results only slightly,  
because such gaps are small \cite{Alford99}.  Note, 
however, that we have removed the degeneracy between $\Delta_{us}$ and 
$\Delta_{ds}$ in order to explore phases in which  these gaps may not be  
equal. 
 
The quasiparticle energies $\lambda_i$ may be obtained by 
diagonalizing the inverse propagator. Equivalently, $\lambda_i$ are 
the eigenvalues of the (72 $\times$ 72) matrix 
\begin{eqnarray} 
D & = &  
\left[ 
\begin{array}{cc} 
- \gamma^0 \vec{\gamma} \cdot \vec{p} - M_{i} \gamma^0 + \mu_{i \alpha} &  
\Delta i \gamma^0 \gamma_5 C \\ 
\gamma^0 C i \gamma_5 \Delta & 
- \gamma^0 \vec{\gamma}^T \cdot \vec{p} + M_{i} \gamma^0 - \mu_{i \alpha}  
\end{array} 
\right] \,, 
\nonumber 
\end{eqnarray} 
where $M_i$ are the dynamically generated quark masses and  $\Delta$ is  
given by 
\begin{equation} 
\Delta = \Delta_{ud}  
\varepsilon^{3 i j} \varepsilon^{3 \alpha \beta} 
+ \Delta_{us} 
\varepsilon^{2 i j} \varepsilon^{2 \alpha \beta} 
+ \Delta_{ds}  
\varepsilon^{1 i j} \varepsilon^{1 \alpha \beta} \, .  
\label{Delta} 
\end{equation} 
\ni Equations (\ref{Lagr}) through (\ref{Delta}) enable a consistent 
model calculation of the thermodynamics of superconducting quark 
matter as a function of the chemical potentials $\mu,~\mu_Q,~\mu_3$, 
and $\mu_8$ at arbitrary temperatures.  For a given set of these 
chemical potentials, the dynamical (or constituent-like) masses $M_i$ and 
the gaps $\Delta_{ij}$ are determined by the solutions of equations 
that result from extremizing $\Omega$ with respect to 
$\langle \bar q_i q_i \rangle$ and $\Delta_{ij}$, respectively. 
 
The phases are labeled according to which of the three gaps are 
non-zero: (1) Normal phase: all gaps zero, (2) 2SC phase: only 
$\Delta_{ud}$ is non-zero, and (3) CFL phase: all gaps are non-zero. 
Where needed, we add electrons simply by noting that $\mu_e=-\mu_Q$, 
and include their free Fermi gas contribution to $\Omega$. 
 
Recall that the ground state is $\tilde{Q}$ neutral, i.e., 
$n_{\tilde{Q}}=-\partial \Omega/ \partial \mu_{\tilde{Q}}=0$, which 
is a consequence of the fact that the condensates are $\tilde{Q}$ 
neutral. Quasiparticles carrying $\tilde{Q}$ charge are massive with 
$m \sim \Delta$ . In addition, the $\tilde{Q}$ susceptibility 
$\chi_{\tilde{Q}}=\partial n_{\tilde{Q}}/\partial 
\mu_{\tilde{Q}}\simeq 0$; in fact, the free energy density is 
independent of $\mu_{\tilde{Q}}$ at zero temperature. This is because 
to generate $\tilde{Q}$ charge in the ground state, we must break a 
pair and the energy cost is of $O[\Delta]$.  In contrast, the free 
energy density depends on $\mu_Q, \mu_3$, and $\mu_8$, and, the 
corresponding individual susceptibilities do not vanish.  For a 
physical $m_s$ of order 100 MeV, there is no apriori reason to expect 
equal numbers of $u,~ d,~{\rm and}~s$ quarks in the CFL phase.  The 
pairing ansatz in Eq.~(\ref{cfldelta}) and the arguments of Rajagopal 
and Wilczek \cite{Rajagopal:2000ff}, however, guarantee that 
\be  
n_{rd}=n_{gu}\,, \qquad n_{bd}=n_{gs}\,, \qquad {\rm and}  
\qquad n_{rs}=n_{bu} \,, 
\ee  
\ni or equivalently, that   
\be  
n_u=n_r\,, \qquad n_d=n_g\,, \qquad {\rm and} \qquad n_s=n_b \,,  
\label{cflscheme} 
\ee 
\ni where $n_{\alpha i}$ is the number density of quarks with color 
$\alpha$ and flavor $i$, and $n_\alpha$ ($n_i$) is the net number 
density of color $\alpha$ (flavor $i$).  Pairing by itself does not 
enforce either color or electric charge neutrality. The strange 
quark mass induces both color and electric charge in the CFL 
phase. We are, however, at liberty to adjust the chemical 
potentials $\mu_3$ and $\mu_8$ to enforce color neutrality. Moreover, 
since the pairing ansatz enforces $\tilde{Q}$ neutrality, enforcing 
color neutrality automatically enforces electric charge neutrality at 
$\mu_Q=0$.  In contrast, the 2SC phase requires a finite $\mu_Q$ to 
satisfy electric charge neutrality and hence admits electrons. 
\begin{figure}[hbt] 
\begin{center} 
\includegraphics[scale=0.4]{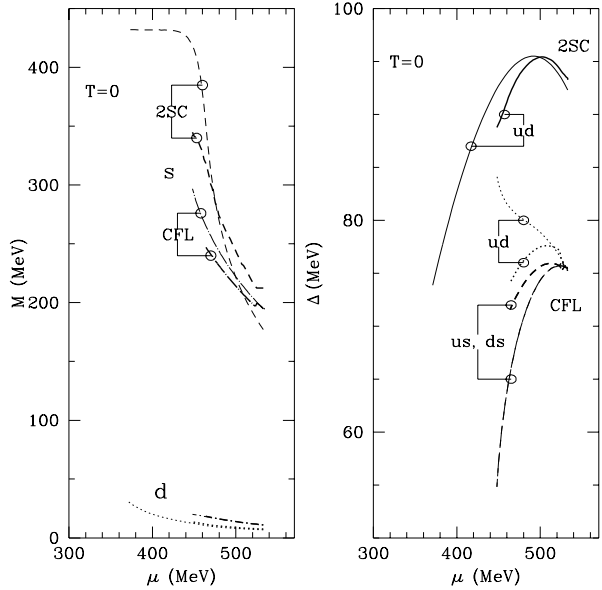} 
\end{center} 
\caption{ Dynamically generated masses and pairing gaps in the CFL 
and 2SC phases at zero temperature from NJL model calculations.  Dark 
(light) curves refer to results when color and electric charge (C \& Q) 
neutrality is (is not) imposed.  } 
\label{fig1} 
\end{figure} 
 
We turn now to discuss results, beginning with those at temperature 
$T=0$.  In Figure~\ref{fig1}, we show the dynamically generated or constituent 
$d$ and $s$ quark masses $M_i$ (left panel) and the pairing gaps 
$\Delta_{ij}$ (right panel) as functions of $\mu$ in the CFL and 2SC 
phases.  The $u$ quark mass, which tracks the trend of the $d$ quark, 
is not shown for the sake of clarity.  The dark (light) curves refer 
to the case in which color and electric charge neutrality is (is not) 
imposed. All masses decrease with increasing $\mu$, since all of the 
$\langle {\bar q}q\rangle$ condensates decrease with $\mu$.  Note that 
the requirement of color and charge neutrality has a larger effect on 
the $s$ quark mass in the 2SC phase than in the CFL phase.  This is 
because neutrality in the 2SC phase requires a large and negative 
electric charge chemical potential. In the discussion that follows, we 
will show that $\mu_Q \sim -m_s^2/2\mu$ in the 2SC phase. Further, 
since $\mu_s = \mu - \mu_Q/3$, a large and negative $\mu_Q$ enhances 
the strange quark density which in turn suppresses the $\langle {\bar 
s}s\rangle$ condensate.  
 
The right panel of Figure~\ref{fig1} shows the various gaps in the CFL 
and 2SC phases. Imposing color neutrality reduces $\Delta_{ud}$, since 
the numbers of red and green quarks (equivalently of $u$ and $d$ 
quarks) are reduced relative to the colored case (see the analytical 
analysis below).  For the same reason, color neutrality increases the 
gaps involving strange quarks.  These trends are broken only when 
$\mu$ begins to approach the ultra-violet cutoff $\Lambda$. The strong 
increase of the $\Delta_{ud}$ gap as $\mu$ decreases for matter in 
which neutrality is not imposed is due to the strong decrease in the 
gaps involving strange quarks.  
\begin{figure}[hbt] 
\begin{center} 
\includegraphics[scale=0.4]{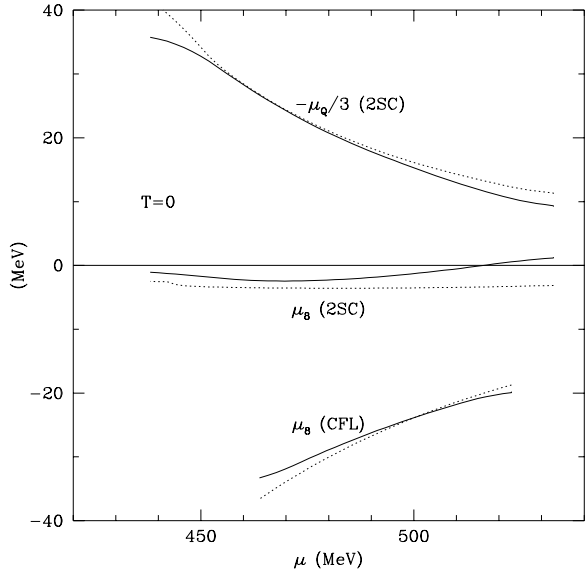} 
\end{center} 
\caption{Chemical potentials $\mu_8$ and $\mu_Q$ that ensure color and 
electric charge neutrality in the CFL and 2SC phases as functions of 
the quark number chemical potential $\mu$ at temperature $T=0$.  Solid 
(dashed) curves refer to results of the NJL (simplified) model. } 
\label{fig2}  
\end{figure} 
 
\begin{figure}[hbt] 
\begin{center} 
\includegraphics[scale=0.4]{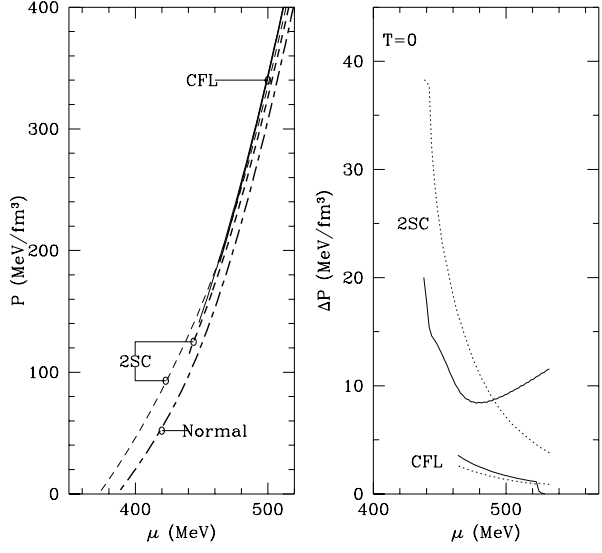} 
\end{center} 
\caption{Left panel: The pressure $P$ versus quark number chemical 
potential $\mu$ in the CFL, 2SC, and normal phases at temperature 
$T=0$. Dark (light) curves refer to the case in which color and 
electric charge (C \& Q) neutrality is (is not) imposed.  Right panel: 
Pressure differences $\Delta P$ or the free energy density cost 
required to ensure C \& Q neutrality in the CFL and 2SC phases at 
$T=0$. Solid (dotted) curves are results of the NJL (simplified) model 
calculations. } 
\label{fig3} 
\end{figure}

In Figure~\ref{fig2}, we show the chemical potentials $\mu_8$ and $\mu_Q$ (as 
functions of $\mu$) required to achieve color and electric charge 
neutrality in the CFL and 2SC phases.  The solid curves refer to 
results of the NJL model calculations.  The left panel of Figure~\ref{fig3} 
shows the pressure $P$ versus $\mu$ at $T=0$. Here the dark (light) 
curves refer to the case in which color and electric charge 
neutrality is (is not) imposed. Note that the pressure of the color 
and electrically neutral normal phase falls below that of the 2SC phase 
for all $\mu$s shown.  The pressure differences $\Delta P$ or the 
free energy density cost necessary to ensure color and electric 
charge neutrality in the CFL and 2SC phases are shown in the right 
panel of Figure~\ref{fig3}.  Here also the solid curves refer to results of 
the NJL model calculations. 
 
In order to gain a qualitative understanding of the results in 
Figures~\ref{fig2} and \ref{fig3}, we undertake an analytical analysis 
of a simpler model also considered in Ref.~\cite{Alford:2002kj}. In 
this analysis, we consider $u$ and $d$ quarks as massless, and include 
corrections due to the $s$ quark mass $m_s$ at leading order as a 
shift in its chemical potential. This does not properly account for 
the shift in energy due the strange quark mass for states far away 
from the Fermi surface, but is a consistent approximation in this 
context since we are primarily interested in the leading order cost of 
enforcing neutrality. Further, we assume that all gaps, including 
those involving the $s$ quark, are independent of 
$m_s$~\footnote{Corrections to $\Delta$ due to $m_s$ arise at 
O[$m_s^2/\mu$]; for a detailed discussion, see Ref. \cite{Kundu:2001tt}.}, 
and that both the gaps and $m_s$ are weak functions of the chemical 
potentials. With these assumptions, and to leading order in $\Delta$ 
\begin{eqnarray} 
\Omega_{\rm CFL} &=& \Omega_{\rm rgb} + \Omega_{\rm rg} + \Omega_{\rm rs} 
                    +  \Omega_{\rm gs} \,, \nonumber \\ 
\Omega_{\rm rgb} &=& -\frac{1}{12 \pi^2} \left( {\mu_{\rm 1}^4} + 
 {\mu_{\rm 2}^4} +{\mu_{\rm 3}^4} + 
{3\Delta^2 (\mu_{\rm 1}^2 +\mu_{\rm 2}^2+ 4 \mu_{\rm 3}^2) }\right)\,, \qquad  
\Omega_{\rm rg} = -\frac{1}{6 \pi^2} \left( {\mu_{\rm rg}^4} +  
{3 \Delta^2 \mu_{\rm rg}^2}\right)  \,, \nonumber \\ 
\Omega_{\rm rb} &=& -\frac{1}{6 \pi^2} \left( {\mu_{\rm rb}^4} +  
{3 \Delta^2 \mu_{\rm rb}^2}\right) \,, \qquad {\rm and} \qquad        
\Omega_{\rm gb} = -\frac{1}{6 \pi^2} \left( {\mu_{\rm gb}^4} +  
{3 \Delta^2 \mu_{\rm gb}^2} \right) \,,  
\label{Omega_CFL} 
\end{eqnarray} 
\ni where we have written the free energy of the CFL phase in terms of 
the $3\times 3$ block involving $ru-gd-bs$ quarks, and three $2\times 
2$ blocks involving $rd-gu$, $rs-bu$ and $gs-bd$ quarks, 
respectively. Each of the three $2\times 2$ blocks is rigid in the 
sense that the free energy is unaffected by differences in chemical 
potentials of quarks that pair in a given  
block~\cite{Rajagopal:2000ff}. The free energy depends only on the 
average chemical potential. The $3\times 3$ block does not exhibit 
this rigidity; here, the quasi-particle energies depend on the splitting 
between the chemical potential characterizing the $ru$, $gd$ and $bs$ 
quarks. Chemical potentials that characterize the free energy of the 
$3\times 3$ block are given by 
\begin{eqnarray} 
\mu_{\rm 1} = \mu + \frac{\mu_8}{\sqrt{3}}  \,, \qquad 
\mu_{\rm 2} = \mu - \frac{\mu_8}{\sqrt{3}} - \frac{m_s^2}{3\mu} \,, \qquad 
{\rm and} \qquad  
\mu_{\rm 3} = \mu - \frac{m_s^2}{6\mu} \,, 
\label{mu3} 
\end{eqnarray} 
\ni and the common chemical potentials that appear in the free energy 
expressions of the $2\times 2$ blocks are given by 
\begin{eqnarray} 
\mu_{\rm rg} = \mu + \frac{\mu_8}{\sqrt{3}}\,, \qquad 
\mu_{\rm rb} = \mu - \frac{\mu_8}{2\sqrt{3}}-\frac{m_s^2}{4\mu}\,,  
\qquad {\rm and} \qquad  
\mu_{\rm gb} = \mu - \frac{\mu_8}{2\sqrt{3}}-\frac{m_s^2}{4\mu} \,. 
\label{mu2common} 
\end{eqnarray} 
\ni We note that, in general, pairing between particles with 
dissimilar masses does not require a common chemical 
potential. Maximal BCS-like pairing requires that the distribution of 
the pairing partners be identical in momentum space.  Since we treat 
the $u$ and $d$ quarks as massless particles and account for the 
effects of the $s$ quark mass through a shift in the chemical 
potential in our analytic analysis, a common chemical potential within 
each pairing block ensures that the aforementioned pairing criterion 
is satisfied. 
 
In the CFL phase, the stress induced by the strange quark mass 
generates color charges. In the limit of nearly equal and vanishing 
light quark masses, the CFL scheme in Eq. (\ref{cflscheme}) indicates 
that we will require only a non-zero $\mu_8$ to achieve color 
neutrality. This justifies why we neglect $\mu_Q$ and $\mu_3$ in 
Eqs.~(\ref{mu3}) and (\ref{mu2common}). To leading order in the parameter 
$m_s^2/\mu$, and assuming that the differences between the various 
gaps are small and $\mu,m_s$-independent, we find that 
 
\be \mu_8 (CFL) = 
-\frac{1}{2\sqrt{3}} \frac{m_s^2}{\mu}~ 
+~O\left[\frac{m_s^4}{\mu^3},\frac{m_s^2 \Delta^2}{\mu^3} \right] 
\label{mu8_CFL} 
\ee  
 
\ni by requiring $\partial \Omega_{CFL}/\partial \mu_8=0$. Since 
$n_u=n_d$ and hence $n_r=n_g$ when there are no electrons, 
$\mu_3(CFL)=0$ identically in the CFL phase at zero 
temperature. Naively, Eq. (\ref{mu8_CFL}) would imply that the free 
energy density cost of enforcing color neutrality in the CFL phase is of 
$O[m_s^2 \mu^2]$.  However, we find that such contributions are absent 
due to cancellations.  This result (see the lower-most dotted curve in 
Figure~\ref{fig2}), with $m_s$ and $\Delta$ of the NJL model as inputs, 
provides an excellent approximation to the exact NJL result.  
Utilizing Eq.~(\ref{mu8_CFL}), we find an analytic estimate for the 
free energy density cost in the CFL phase: 
\be 
\Delta\Omega_{CFL} =  \Omega_{CFL}(\mu_8) - \Omega_{CFL}(0)   
=  \frac{5 m_s^4}{72 \pi^2}  
+~O\left[\frac{m^6}{\mu^2},\frac{m^4\Delta^2}{\mu^2}\right] \,.  
\label{ecost_CFL} 
\ee The lower dotted curve in the right panel of Figure~\ref{fig3} shows that 
this result is in quantitative agreement with the NJL model 
calculation.  
 
In the 2SC phase, the pairing phenomenon itself gives rise to color 
charges.  The 2SC thermodynamic potential, to leading order in the gap 
and consistent with the approximation scheme described earlier, is 
\begin{eqnarray} 
\Omega_{2SC} &=& \Omega_{\rm rugd} + \Omega_{\rm free} \,, \nonumber \\ 
\Omega_{\rm rugd} &=& -\frac{1}{3 \pi^2} \left({\mu_{\rm rugd}^4}  
 +  {3 \Delta^2 \mu_{\rm rugd}^2}\right) \,, \nonumber \\ 
\Omega_{\rm free}&=& -\frac{1}{12 \pi^2} \left( {\mu_{\rm bu}^4} 
 +{\mu_{\rm bd}^4} 
 +{\mu_{\rm bs}^4} 
 +{\mu_{\rm rs}^4} 
 +{\mu_{\rm gs}^4} 
 +{\mu_Q^4}\right) ~\,.  
\label{Omega_2SC} 
\end{eqnarray} 
\ni The chemical potentials appearing above are defined by 
\begin{eqnarray} 
\mu_{\rm rugd} &=& \mu + \frac{\mu_8}{\sqrt{3}} + \frac{\mu_Q}{6}\,, \qquad 
\mu_{\rm bu} = \mu - \frac{2\mu_8}{\sqrt{3}} + \frac{2\mu_Q}{3}\,, \qquad 
\mu_{\rm bd} = \mu - \frac{2\mu_8}{\sqrt{3}} - \frac{\mu_Q}{3}\,,\nonumber \\ 
\mu_{\rm bs} &=& \mu - \frac{2\mu_8}{\sqrt{3}} - \frac{\mu_Q}{3}  
- \frac{m_s^2}{2\mu}\,, \qquad {\rm and} \qquad  
\mu_{\rm rs} = \mu_{\rm gs} = \mu + \frac{\mu_8}{\sqrt{3}}  
- \frac{\mu_Q}{3} - \frac{m_s^2}{2\mu} \,. 
\end{eqnarray} 
 
\ni The condition to ensure color neutrality, $\partial 
\Omega_{2SC}/\partial \mu_8=0$, yields 
\be  
\mu_8 (2SC) = 
-\frac{1}{3\sqrt{3}} \frac{\Delta^2}{\mu}  
+~O\left[\frac{\Delta^4}{\mu^3}\right] \,,  
\label{mu8_2SC}   
\ee  
\ni where we have used a common value of $\Delta$ (independent of 
$\mu$) in the analytical analysis.  Note that $\mu_8 (2SC)$ does not 
depend on $m_s$ at leading order.  Since pairing in the 2SC phase involves  
red and green quarks, it does not induce a color 3-charge; hence  
$\mu_3(2SC)=0$.  However, electric charge neutrality in the 
2SC phase requires an adjustment due to the magnitude of $m_s$.  At 
leading order in a $1/ \mu$ expansion, we find 
 
\be  
\mu_Q (2SC) = -\frac{1}{2} \frac{m_s^2}{\mu} -\frac{1}{3} 
\frac{\Delta^2}{\mu} +~O\left[\frac{m_s^4}{\mu^3},\frac{\Delta^2 m_s^2}{\mu^3} 
\right]  
\label{muQ_2SC}   
\ee  
 
\ni by setting $\partial \Omega_{2SC}/\partial \mu_Q=0$.  As in the 
CFL phase, the free energy density cost of enforcing color neutrality 
in the 2SC phase is small, because $O[\Delta^2 \mu^2]$ terms cancel 
and the free energy density begins to change at 
$O[\Delta^4]$. Similarly, we find that there is no cost for enforcing 
electric charge neutrality in the 2SC phase at $O[\mu^2]$. Using the 
results in Eqs.~(\ref{mu8_2SC}) and (\ref{muQ_2SC}), the free energy 
density cost of enforcing color and electric charge neutrality becomes 
\be \Delta \Omega_{2SC} = \Omega_{2SC} (\mu_8,\mu_Q) - \Omega_{2SC} 
(0,0) = \frac{1}{8 \pi^2} \left ( m_s^4 + \frac{4\Delta^2m_s^2}{3} + 
\frac{4\Delta^4}{3} \right) +O\left[\frac{m_s^6}{\mu^2}, 
\frac{\Delta^6}{\mu^2}, \frac{\Delta^2 m_s^4}{\mu^2}\right] \,. 
\label{ecost_2SC} 
\ee 
Although the free energy density costs of enforcing color neutrality in the 
CFL and 2SC phases are of the same order, the cost in the 2SC phase is 
numerically larger. This is in part due to the larger strange quark 
mass in the 2SC phase and because the free energy density cost due to $\Delta$ 
and $m_s$ dependent terms add in the 2SC phase.  The analytical 
results in Eqs.~(\ref{mu8_2SC}), (\ref{muQ_2SC}), and 
(\ref{ecost_2SC}), shown as dotted curves in Figures~\ref{fig2} and  
\ref{fig3}, compare well with the results of the NJL model.  
 
\ni {\bf Phase diagram at finite temperature and lepton content:} In 
the proto-neutron star context, matter is subject to stresses induced 
by finite temperature and lepton number chemical 
potentials~\cite{prakash:1997}. Since electrons have both electric and 
lepton number charges,  
\be  
\mu_e = - \mu_Q + \mu_{L e} \,, 
\ee  
\ni where $\mu_{L e} = \mu_{\nu_e}$ is the chemical potential for 
electron lepton number.  In order to explore the effects of a finite 
neutrino chemical potential and finite temperature on the 
superconducting phases, we employ the NJL model, Eq.~(\ref{Omega1}) 
with extensions to include neutrinos and electrons. 
\begin{figure}[hbt] 
\begin{center} 
\includegraphics[scale=0.4]{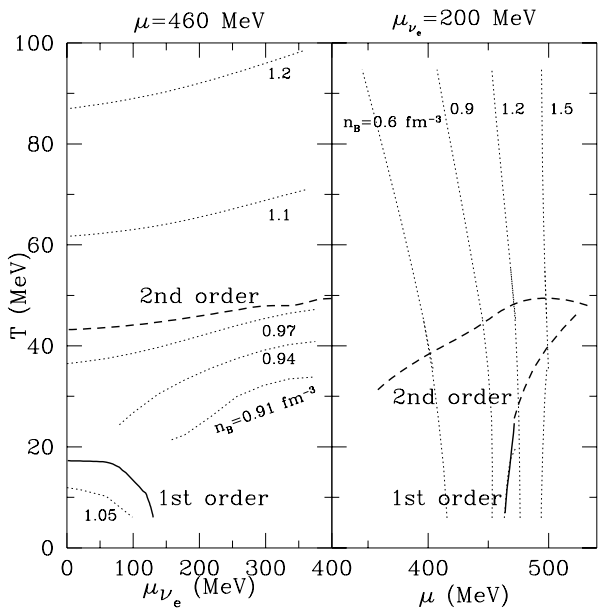} 
\end{center} 
\caption{Cross-sectional views of the $T-\mu - \mu_{\nu_e}$ phase 
diagram at the indicated values of $\mu$ and $\mu_{\nu_e}$. In both 
panels, the dark curves show the phase boundaries, while the dotted 
curves show contours of constant baryon density. } 
\label{fig4} 
\end{figure} 
 
Figure~\ref{fig4} shows representative cross-sectional views of the 
$T-\mu - \mu_{\nu_e}$ phase diagram. The left panel displays results 
at fixed $\mu=460$ MeV. With increasing temperature, a first order 
transition occurs from the CFL phase to the 2SC phase with $ud$ 
pairing.  For $\mu_{Le}=0$, the transition occurs at $T \simeq 17$ 
MeV.  The corresponding baryon densities are $n_B=(n_u+n_d+n_s)/3 
=1.06~{\rm fm^{-3}}$ in the CFL phase and $n_B=0.94~{\rm fm^{-3}}$ in 
the 2SC phase.  At zero neutrino chemical potential, an analytic 
estimate of the critical temperature $T_c$ for the CFL-2SC transition 
can be obtained by assuming that the gaps in the 2SC and CFL phases 
are nearly equal to each other and to their zero temperature values.  
We find that 
 
\begin{eqnarray}  
T_c &=& \frac{\sqrt{2}}{\pi}\Delta_0~\left[1 - \frac{4}{5} 
\frac{\delta \Delta}{\Delta_0}~\xi_{\rm BCS}^2   
+ \frac{9}{20} \frac{m_s^2}{\Delta_0^2} \frac{\delta m_s}{m_s} \right]~ 
\xi_{\rm BCS} \qquad {\rm with} \nonumber \\ 
\xi_{\rm BCS} &=& \left(1 + 2 \frac{\Delta_0^2}{\pi^2 T_{\rm BCS}^2} 
\right)^{-1} \,,  
\label{tccfl2sc} 
\end{eqnarray} 
  
\ni where $\Delta_0$ is the zero temperature gap, $\delta \Delta$ and 
$\delta m_s$ are the differences between the gaps and the strange 
quark masses in the 2SC and CFL phases, respectively. $T_{\rm BCS}$ is 
the temperature at which the gap in the CFL phase would vanish, 
assuming that $\Delta(T) = \Delta_0 \sqrt{1-(T/T_{\rm BCS})^2}$. Note 
that, in general, gaps involving strange quarks do not vanish at the 
transition, i.e, the phase transition is first order.  This is because 
at leading order, the critical temperature $T_c \sim \Delta_0/2.22$ 
is less than that for the second order BCS transition, $T_{\rm 
BCS}\simeq \Delta_0/1.76$. It is clear from Eq. (\ref{tccfl2sc}) 
that contributions to $T_c$ due to $\delta \Delta$ and $\delta m_s$ 
can easily alter this, allowing for a BCS like second order 
transition. If the magnitude of the gap in the 2SC phase is larger 
than that in the CFL phase, $T_c$ is lowered and the transition 
becomes more strongly first order.

Accommodating a finite lepton number in the CFL phase is expensive, 
because the requirement of color and electric charge neutrality in 
this phase excludes electrons.  At $T=0$, the transition from the CFL 
to 2SC phase occurs at $\mu_{Le} \simeq 150$ MeV.  The latent heat 
density, $T \Delta (\partial P / \partial T$), lies in the range 
(2--15) MeV/fm$^3$ along the boundary of the first order phase 
transition.  With increasing temperature, the critical lepton chemical 
potential at which the CFL-2SC transition occurs decreases.  This is 
because the gaps in the CFL phase decrease with increasing $T$; 
consequently, unlocking occurs at smaller $\mu_{\nu_e}$.

The right panel in Figure~\ref{fig4} shows the phase boundaries at 
fixed $\mu_{\nu_e}=200$ MeV. For this neutrino chemical potential, the 
CFL phase is preferred above $\mu=460$ MeV.  For low (high) values of 
$\mu$, the region of the CFL phase shrinks (expands) progressively to 
lower (higher) values of $T$ and $\mu_{\nu_e}$. In contrast, the 
2SC-Normal phase boundary is relatively unaffected by increasing 
values of $\mu$ (in the range relevant for proto-neutron star 
studies), although minor variations do occur. Note that with 
increasing temperature, the phase transition switches from first to 
second order. This switch is due to the fact that the $\Delta_{us}$ 
and $\Delta_{ds}$ gaps decrease along the first order phase transition 
line. When these gaps vanish (at $T\sim 25$ MeV for $\mu_{\nu_e}=200$ 
MeV), the phase transition becomes second order.  In 
Figure~\ref{fig4}, contours of constant baryon density are shown by 
the dotted curves in both panels.  Notice that, for the values of 
$\mu$ and $\mu_{\nu_e}$ chosen for display, the 2SC phase supports 
lower baryon densities than the CFL phase. Across the first order 
phase transition, the density contours are discontinuous.  We wish to 
add that the phase in which $\Delta_{us} \neq \Delta_{ds}$ was found 
to be thermodynamically disfavored in the range of $T,~\mu,~{\rm 
and}~\mu_{\nu_e}$ explored here. 
 
The consequences of requiring local color neutrality in 
superconducting quark matter with and without neutrinos at both zero 
and finite temperatures are the principal findings of this work. 
Quantitative results, especially those for quark number chemical 
potentials approaching the ultra-violet cut-off in the NJL model used, 
should be viewed with some caution.  Notwithstanding this, the basic 
qualitative features concerning the phase transitions appear to be 
generic, insofar as similar trends are found in our analytic analysis 
that employed a simplified model without a cutoff.  We also wish to 
emphasize that the phase diagram in Figure~\ref{fig4} requires 
revision at low values of $\mu$ (or low baryon densities) for which a 
hadronic phase is more likely to be favored.

\ni {\bf Kaon condensation:}  
In neutrino-free matter, Bedaque and Schafer have shown that the 
strange quark mass induces a stress which can result in the 
condensation of neutral kaons in the superconducting quark phase 
\cite{Bedaque:2001je,Kaplan:2001qk}.  Kaon condensation occurs when 
the stress induced by the strange quark mass $m_s^2/(2 \mu) \gsim 
m_{K^0}$, where $m_{K^0}$ is the mass of the neutral kaon.  In 
general, the masses of all pseudo-Goldstone bosons receive 
contributions both from the diquark and quark-antiquark 
condensates. For example, the mass of the neutral kaon is given by  
\cite{Son:1999cm,Manuel:2000wm,Schafer:2001za,Schafer:2002ty} 
 
\be m_{K^0}^2= a~m_u(m_d+m_s) + \chi (m_d+m_s)\,, \ee where $a \sim 
\Delta^2/\mu^2$ and $ \chi \sim \langle \bar{q} q\rangle \,$.  At 
asymptotically high density, where the axial $U(1)$ symmetry is 
restored and the quark-antiquark condensate vanishes, the dominant 
contribution to the masses is from the diquark condensate. In this 
case, the kaon masses become small, of order 10 MeV, and kaon 
condensation is robust. At the densities of relevance to neutron 
stars, the situation is less clear because a finite $\langle \bar{q} 
q\rangle$ can potentially result in larger masses for the kaons and 
disfavor meson condensation 
\cite{Manuel:2000wm,Schafer:2002ty}. Pending a detailed investigation 
of this question within the NJL model, we assess here how kaon 
condensation can affect the structure of the color-neutral phase 
by assuming that the kaon mass is small compared to $m_s^2/(2 
\mu)$. This corresponds to near maximal kaon condensation. In this 
case, the meson contribution to the pressure is easily computed and is 
given by \cite{Kaplan:2001qk} 
 
\be 
P_{K^0} = \frac{1}{2} f_{\pi}^2~\left[\frac{m_s^2}{2 \mu}\right]^2~ 
\left(1 + O\left[ \frac{m_{K^0}^2 \mu^2 }{m_s^4} \right] \right) \,, 
\ee 
where $f_{\pi} \sim \mu $ is the pion decay constant in the effective 
theory describing Goldstone bosons \cite{Son:1999cm}. The leading 
contribution to the pressure from the Goldstone bosons is of order 
$m_s^4$. Including this contribution we find that at 
zero temperature and neutrino chemical potential, the phase transition 
between the 2SC and CFLK$0$ phases occurs at a value of $\mu$ which is 
lower by $16$ MeV compared to the case without kaons. 
 
At finite neutrino chemical potential, CFL quark matter contains a 
K$^+$ condensate which admits electrons, even at zero temperature 
\cite{Kaplan:2001qk}. In the presence of electrons, charged kaon 
condensation lowers the free energy density cost for accommodating 
lepton number. Including the electron and $K^+$ contributions to the 
pressure ($P_{K^+}=f_{\pi}^2 \mu_{K^+}^2/2$), and solving 
self-consistently for the condition of charge neutrality, the pressure 
is increased by about 2 MeV at $\mu_{\nu_e}=200$ MeV. These results 
indicate that the extent of the meson condensed CFL phase in the phase 
diagram is likely to be enlarged, but only by a few percent. In our 
analysis, the order of the phase transitions between the CFL and the 
2SC phase was not affected by meson condensation. We wish to note, 
however, that our analysis neglects the effect of the meson condensate 
on the quasi-particles themselves. This feedback, which can alter the 
quark contribution to the free energy, must be explored before 
quantitative conclusions regarding the role of meson condensation on 
the phase diagram can be drawn. This warrants further investigation 
and is beyond the scope of this article.

\ni {\bf Astrophysical Implications:} The $T-\mu-\mu_{\nu_e}$ phase 
diagram offers clues about the possible phases encountered by a 
neutron star from its birth as a proto-neutron star (in which 
neutrinos are trapped) in the wake of a supernova explosion to its 
neutrino-poor catalyzed state with ages ranging from hundreds of 
thousands to million years. In earlier work, some aspects of how a 
phase transition from the normal to the 2SC phase would influence 
neutrino transport in a newly-born neutron star were explored 
\cite{Carter:2000xf}. To date, detailed calculations of the 
evolution of a proto-neutron star with quarks have been performed for 
the case in which only the normal phase was considered 
\cite{Pons:2001}.  Our findings in this work indicate that the core of 
a proto-neutron star may well encounter a 2SC phase first when matter 
is hot and neutrino-rich before passing over to a CFL phase.   
 
\ni {\bf Conclusions:} Color and electric charge neutrality in the 
superconducting quark phases requires the introduction of chemical 
potentials for color and electric charge. The magnitudes of these 
chemical potentials are sub-leading in $\mu$. The corresponding free 
energy density costs are small and independent of $\mu$ at leading 
order with the free energy density cost for neutrality in the 2SC 
phase being significantly larger than that in the CFL 
phase. Consequently, and in agreement with Ref. \cite{Alford:2002kj}, 
we find that the bulk 2SC phase is less likely to occur in compact 
stars at $T=0$ and $\mu_{\nu_e}=0$.  In the NJL model, a small 2SC 
window does exist at relatively low baryon density. However, since 
this window occurs at very low density it is likely to be shut by the 
hadronic phase. If homogeneous quark matter were to occur in neutron 
stars, it seems likely that with increasing $\mu$ a sharp interface 
would separate hadronic matter and CFL quark matter 
\cite{Alford:2001zr}. We note, however, that we have only considered 
homogeneous phases in this study and it is possible that less 
symmetric heterogeneous phases may well be favored for chemical 
potentials of relevance to neutron stars.  Examples include the 
CFL-Hadron mixed phase \cite{Alford:2001zr} and crystalline 
superconductivity \cite{Bowers:2002xr}.  These possibilities alleviate 
the cost of enforcing color neutrality, since in these cases it is 
only a global constraint. In such phases, however, energy costs 
associated with gradients in particle densities must be met. 
 
The value of the diquark coupling employed in this work predicts gaps 
on the order of $100$ MeV at $\mu\simeq 500$ MeV. The relationship between 
the $\langle qq \rangle$ and $\langle{\bar q} q\rangle$ condensates in 
medium obtained by employing the mean-field gap equations with
the couplings set by the Fierz transformation in vacuum may differ 
from that obtained in a more exact treatment of the NJL model.
Lacking experimental guidance on their values in medium, we have 
studied the influence of moderate changes to the diquark coupling (but 
within the mean-field approximation) on the predicted phase structure. 
For example, using $G_{DIQ} = G_S$, we find that gaps are increased by 
about 20\% in both the CFL and 2SC phases relative to the case with 
$G_{DIQ} = 3G_S/4$ predicted by the Fierz transformation in vacuum. 
The extent of the low density region in which a 2SC phase is favored 
over the CFL phase is not greatly changed and the 2SC phase continues 
to be favored over the normal phase.

Had we ignored the differences in the strange quark mass between the 
charge-neutral normal phase and the charge-neutral 2SC phase, the 
normal phase would be favored over the 2SC phase at low density when 
$\Delta \lsim m_s^2/4\mu$ \cite{Alford:2002kj}. However, we find that 
the chemical potential for the strange quarks is larger in the 
charge-neutral 2SC phase than that in the normal phase. Consequently, 
$\langle\bar{s} s\rangle$ is reduced and the lighter strange quarks in 
the 2SC phase contribute more to the pressure. It is this feedback 
that tips the balance in favor of the 2SC phase when $\Delta \lsim 
m_s^2/4\mu$.  However, we do not expect this trend to remain intact 
for larger variations in the couplings. Obviously, it is always 
possible to reduce $G_{DIQ}$ and increase $G_S$ so as to allow for the 
existence of a normal phase at lower density.  At finite temperature 
and neutrino chemical potential, the CFL phase becomes less favored 
both because of its small specific heat and because of its 
exponentially suppressed (by the factor $\exp({-\Delta/T}))$ electron 
number density, which makes the free energy density cost of 
accommodating lepton number large. In contrast, the 2SC phase has a 
larger specific heat and easily accommodates electron number, and is 
therefore the favored phase at finite temperature and lepton number.

The inclusion of Goldstone bosons in the CFL phase tends to extend the 
region in the $T-\mu_{\nu_e}$ plane where the CFL phase is favored, 
since Goldstone bosons contribute significantly to the specific heat 
and also allow for the presence of electrons. In the absence of 
Goldstone bosons, a first order unlocking transition occurs from the 
CFL phase to the less symmetric 2SC phase with increasing lepton 
chemical potential.  When the temperature is sufficiently high, the 
phase transition switches from first to second order due to the fact 
that the $\Delta_{us}$ and $\Delta_{ds}$ gaps decrease along the first 
order phase transition line and eventually vanish. 
 
As discussed above, different phases of color superconducting 
quark matter are likely to be traversed by the inner core of a 
proto-neutron star during its early thermal evolution. The task ahead 
is to study how these phases and transitions between them influence 
observable aspects of core collapse supernova, neutron star structure, 
and thermal evolution. 
 
%%%%%%%%%%%%%%%%%% 
% 
%%%%%%%%%%%%%%%%%% 
 
\vskip 0.25in  
 
The work of A.W.S. and M.P. was supported by the U.S. Department of 
Energy grant DOE/DE-FG02-87ER-40317 and that of S.R. was supported in 
part by the U.S. Department of Energy under the cooperative research 
agreement DF-FC02-94ER40818.  We thank Mark Alford, Krishna Rajagopal, 
Prashanth Jaikumar, James Lattimer, and Thomas Sch\"afer for several useful 
discussions. 
 
\nopagebreak

\end{document}